\documentclass[conference]{IEEEtran}
\usepackage[utf8]{inputenc}
\usepackage{cite}
\usepackage{amsmath,amssymb,amsfonts}
\usepackage{algorithmic}
\usepackage{graphicx}
\usepackage{textcomp}
\usepackage[table,xcdraw]{xcolor}
\usepackage{subcaption}
\usepackage{tcolorbox}
\usepackage[capitalize]{cleveref}
\usepackage[acronym]{glossaries}
\usepackage{svg}
\usepackage{makecell}
\makeglossaries
\newacronym{ap}{AP}{Access Point}
\newacronym{ai}{AI}{Artificial Intelligence}
\newacronym{gearr}{GEARR}{goal-effective achievable rate region}
\newacronym{ber}{BER}{Bit Error Rate}
\newacronym{go}{GO}{goal-oriented}
\newacronym{do}{DO}{data-oriented}
\newacronym{kpi}{KPI}{Key performance Indicator}
\newacronym{ml}{ML}{Machine Learning}
\newacronym{es}{ES}{edge server}
\newacronym{noma}{NOMA}{non-orthogonal multiple access}
\IEEEoverridecommandlockouts
\usepackage{caption}
\usepackage{textcomp}
\usepackage{dsfont}
\usepackage{balance}
\usepackage{subcaption}
\usepackage{enumitem}
\usepackage{titlesec}
\newtheorem{definition}{Definition}

\title{Goal-oriented Spectrum Sharing: Trading Edge Inference Power for Data Streaming Performance\vspace{-0.2 cm} \thanks{This work has been supported by the SNS JU project 6G-GOALS under the EU’s Horizon program Grant Agreement No 101139232, and by the ANR under the France 2030 program, grant "NF-NAI: ANR-22-PEFT-0003". For Miltiadis C. Filippou this work initiated while being with Nokia Strategy $\&$ Technology, 81541 Munich, Germany. Fig. \ref{fig:sys_model} has been partly designed using resources from Flaticon.com}}
\author{Mattia Merluzzi$^1$ and Miltiadis C. Filippou$^2$\\
$^1$CEA-Leti, Université Grenoble Alpes, F-38000 Grenoble, France\\
$^2$WINGS ICT Solutions, 17121 Athens, Greece \smallskip\\
e-mail: mattia.merluzzi@cea.fr,  mfilippou@wings-ict-solutions.eu \vspace{-0.5 cm}}

\begin{document}
\maketitle

\begin{abstract}
We study the problem of spectrum sharing between goal-oriented (GO) and legacy data-oriented (DO) systems. For the former, data quality and representation is no longer optimized based on classical communication key performance indicators, but rather configured on the fly to achieve the goal of communication with the least resource overhead. This paradigm can be followed to flexibly adapt wireless and in-network artificial intelligence operations across different nodes (e.g., access points, users, sensors or actuators) to data traffic, channel conditions, energy availability and distributed computing capabilities. In this paper, we argue and demonstrate that computing and learning/inference operation performance strongly affect lower layers, calling for a real cross-layer optimization that encompasses physical and computation resource orchestration, up to the application level. %We propose to introduce this cross-domain flexibility toward solving the coexistence problem of two systems sharing the radio spectrum, namely, a goal-oriented and a legacy transmission/ reception pair, where one interfers with the other. We argue and demonstrate that computing and learning/inference operation approaches and performance strongly affect lower layer network performance, calling for a real cross-layer optimization that encompasses physical layer, compute resource orchestration, and applications. 
Focusing on a communication channel shared among a GO and a DO user, we define a \textit{goal-effective achievable rate region (GEARR),} to assess the maximum data rate attainable by the latter, subject to goal achievement guarantees for the former. Finally, we propose a cross-layer dynamic resource orchestration able to reach the boundaries of the GEARR, under different goal-effectiveness and compute resource consumption constraints. %Our framework opens new opportunities for a goal- and computation-aware spectrum management. %Numerical results show that this flexible cross-layer optimization scheme outperforms benchmark strategies and policies.
\end{abstract}
\begin{IEEEkeywords}
Goal-oriented semantic communications, adaptive computation, resource allocation, spectrum sharing. 
\end{IEEEkeywords}
\section{Introduction}
Semantic and \gls{go} communication aims at dynamically tailoring data representation and transmission, as guided by specific application needs \cite{6G_GOALS}. Within the scope of this promising paradigm for 6G, communication performance requirements are \textit{adapted to achieve the communication goal}, rather than set a priori and ossified. Wireless resource sharing between semantic, \gls{go} and legacy \gls{do} services has several implications on network architectural design, along with radio and computing resource deployment and orchestration. It comes with challenges in terms of system backward compatibility, but also opportunities for more efficient spectrum use, thanks to the extraction of relevant information, also possibly exploiting the much lower application data quality that can be tolerated during transmission for some tasks. An indicative example of such tasks refers to the ones involving advanced \gls{ai}-based processing, including computer vision models dedicated to image classification or object detection that exhibit substantial robustness to noise (or, bit-level errors). Semantic data extraction and processing as part of a \gls{go} communication system setup need computing resources to be capillary available at end devices and edge nodes (e.g., \gls{es}). Such edge resources facilitate the achievement of low two-way latency requirements, energy consumption reduction, data privacy and security, with data being kept as local as possible. Further, these resources assist with extracting relevant information, thus overcoming wireless signaling drawbacks (e.g., bit-level errors). This capability magnifies with increased computing capacity, and represents an opportunity for more efficient spectrum sharing.\\% However, the trade-off shall always be assessed.\\%, however with compute-resource awareness. \\
\noindent\textbf{Related works.} A few works have already focused on spectrum coexistence between semantic, \gls{go} and \gls{do} systems. %However, a blind massive use of resources does not meet the requirements for more sober networks in terms of material manufacturing and energy consumption during operation. Maximizing performance cannot be the only objective, we need rather to shift towards an energy/carbon-aware resource management, with energy availability and sobriety at the center of system orchestration and management. Therefore, the semantic and goal-oriented paradigm is a huge opportunity, but also the source for a fundamental trade-off between communication and computing. 
In \cite{Xidong23}, the authors propose a semi-\gls{noma} scheme for a two-users downlink communication, to improve the achievable rate for a \gls{do} user. %They show how semantic communication can improve performance of a \gls{do} communication in terms of achievable rate region. 
However, the authors focus solely on the conveyed semantics, while overlooking the goal of communication. A similar approach is proposed in \cite{Liu24} for uplink communication, considering different multiple access schemes, to characterize the trade-off between semantic user rate and \gls{do} user rate. Again, the communication goal is limited to correctly receiving message meaning. In \cite{Merluzzimdpi23}, \gls{go} communication is introduced in the problem, with a scheme that proposes to learn an adaptation of goal-achieving communication quality metrics to \gls{do} user interference, to allow a \gls{go} user to achieve its goal, i.e., confident and timely inference. None of these works proposes goal-aware adaptation of computing resources to the quality of received data in case of unfavorable channel conditions.\\
\noindent\textbf{Contribution.} We tackle this heterogeneous service coexistence problem from an interference perspective, to show how radio and computation resource domains are tightly related. Going beyond previous works, we propose to incorporate computing resource awareness and inference model availability into the resource orchestration policy. We define the concept of \textit{\gls{gearr}}, and propose a dynamic method to jointly control \gls{do} user transmit power, inference model selection for the \gls{go} user, proactive packet drops and computation resources, to explore its boundaries.\\ %A long-term formulation is efficiently transformed into a series of instantaneous problems, based on cross-layer instantaneous observations. Finally, numerical results show the capability of the proposed joint methodology to explore the boundaries of the \gls{gearr}s.\\
\textbf{\textit{Notation:}} in the remainder of the paper, bold lower case letters denote vectors, while calligraphic letters denote sets. Also, given a random variable X, its long-term average is always denoted as $\bar{X}$, and defined as
\begin{equation}\label{long_term_avg}
   \bar{X}=\lim\nolimits_{T\to\infty}\frac{1}{T}\sum\nolimits_{t=0}^{T-1}\mathbb{E}\{X(t)\}.
\end{equation}
%On one hand, a computing-resource aware radio resource management can help greatly enhancing performance. On the other hand, computing availability should depend on energy availability, with energy mix being an additional source for carbon-aware resource management. We will show how the performance of legacy users using the radio spectrum shared by a goal-oriented system, strongly depend on the availability of energy, its mix, and the requirements of the goal-oriented service. We consider two cases: i) the \gls{go} and \gls{do} users camp on the coverage area of the same wireless \gls{ap}, and ii) each of the \gls{go} and \gls{do} users communicates to its application server via a different AP, albeit operating on the same channel. Depending on the information that the network and the users share, different performance can be achieved. The objective is to maximize the performance of the \gls{do} user under constraints on: i) the goal achievement of the \gls{go} user and ii) long-term computational load constraints for the system.
\section{System model}\label{sec:sys_model}
The system under investigation is composed of two users, namely a \gls{go} user ($\text{UE}_g$), and a \gls{do} user, ($\text{UE}_d$), both served by the same \gls{ap} in uplink, on the same frequency resources. The \gls{ap} is equipped with $N$ antennas and a computing node that is embedded with a set $\mathcal{L}$ of pre-trained \gls{ai} models, ready to output inference results for $\text{UE}_g$. The system is illustrated in Fig. \ref{fig:sys_model}. The role of the buffer and the \textit{control valve} at $\text{UE}_d$ will be clarified in Sec. \ref{sec:problem_formulation}. Denoting by $\mathbf{h}_{g(d)}(t)\in\mathbb{C}^{N\times 1}$ the complex channel for $\text{UE}_{g(d)}$ at a given time instant $t$, we can write the (instantaneous) Signal-to-Noise-plus-Interference-Ratio ($\text{SINR}_{d(g)}$) as
\begin{equation}\label{sinr}
    \text{SINR}_{g(d)}(t)=\frac{|\mathbf{w}_{g(d)}^H(t) \mathbf{h}_{g(d)}(t)|^2p_{\text{tx},g(d)}(t)}{|\mathbf{w}_{g(d)}^H(t)\mathbf{h}_{d(g)}(t)|^2 p_{\text{tx},d(g)}(t) +\sigma_n^2}, 
\end{equation}
with $p_{\text{tx},g(d)}$ and $\sigma_n^2$ denoting the transmission power of $\text{UE}_g$ ($\text{UE}_d$) and the noise power, respectively; whereas $\mathbf{w}_{g(d)}$ is the \gls{ap} combining vector for the \gls{go}/\gls{do} user. We assume that the \gls{ap} has instantaneous channel knowledge and applies a signal reception technique, e.g., Maximum Ratio Combining (MRC).

%In the next sections, we detail the KPIs related to the \gls{go} and \gls{do} user, both in terms of wireless (delay, \gls{ber}, data rate) and computing aspects.
\subsection{Key Performance Indicators (KPIs)}\label{sec:wireless_perf}
In this paper, we are interested in wireless performance of both the \gls{go} and \gls{do} user. For $\text{UE}_g$, we need to consider the following communication KPIs: i) communication delay and ii) communication reliability. For the latter metric, we use the \gls{ber}, which affects inference performance, as detailed in the sequel. Further, for $\text{UE}_g$, we consider computing delay as impacting inference timeliness. For $\text{UE}_d$ we consider the data rate as the dominant communication KPI to support the running application, which is, however, rather insensitive to the relevance of communicated data. The objective is to assess the performance of the interference channel in terms of \textbf{\textit{goal-effective achievable rate regions}}.
\subsubsection{Goal-oriented user: wireless delay, \gls{ber} and inference}
We assume the \gls{ap} to dynamically select, at each time $t$, a modulation order $M(t)\in\mathcal{M}$ for UE$_g$, with an $M$-QAM constellation. Given $M(t)$, the wireless communication delay to upload a new inference input data batch reads as $D_{\text{tx}}(t) =  N_b(t)/R_g(t)$, 
%\begin{equation}\label{comm_delay}
%    D_{\text{tx}}(t) =  N_b(t)/R_g(t),
%\end{equation}
where $N_b(t)$ is the number of bits encoding one data batch,  $R_g(t)=W\log_2(M(t))$ is the UE$_g$ data rate and $W$ is the available uplink bandwidth. Note that $N_b$ could evolve over time thanks to possibly different source compression schemes that depend on the available resources and the specific context on the fly. However, in this paper we keep it fixed over time. The \gls{ber} $P_b(t)$ depends on SINR$_g(t)$ and $M(t)$, and for an uncoded modulation is given by \cite{goldsmith05}:
\begin{equation*}
    P_b(t) =  \frac{4}{\log_2(M(t))}\left(1-\frac{1}{\sqrt{M(t)}}\right)Q\left(\sqrt{\frac{3\cdot\text{SINR}_g(t)}{M(t)-1}}\right).
\end{equation*}

%The case of coded modulation (modulation and coding scheme selection) can be considered and is left for future work.
\subsubsection{Computing aspects: delay and inference KPIs}
\begin{figure}[t]
    \centering
    \includegraphics[width=0.8\columnwidth]{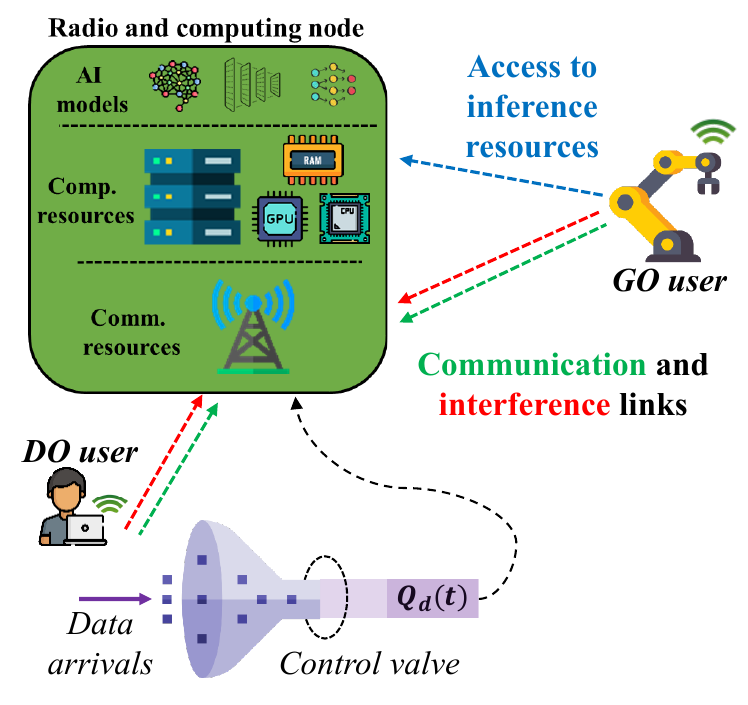}
    \caption{Reference scenario}
    \vspace{-0.5 cm}
    \label{fig:sys_model}
\end{figure}
In this work, computing only concerns the \gls{go} user. %, although the framework could be easily generalized to multiple connect-compute services sharing the same infrastructure. 
Also, we assume the \gls{es} to be embarked with a set of $\mathcal{L}=\{1,\ldots, L\}$ inference models (e.g., \gls{ai} models) capable of addressing the \gls{go} user's task, each one with different computational complexity, thus different performance and robustness to data noise. Every model $l$ is characterized by a tuple $(\omega_l, \Gamma_l(P_b))$, where $\omega_l$ is the number of Floating-Point-Operations (FLOPs) needed to run one inference instance (i.e., inference on one data batch)\footnote{Note that $\omega_l$ is an average value that does not take into account sample-specific computational cost. We leave this for future work.}, %\textcolor{magenta}{[MF]: should we mention that $\omega_l$ is an average value for the l-th model? I am thinking that different inference input may differently fire-up activation functions during a forward pass in inference, thereby translating into more or less additions/ multiplications each time.}, 
and $\Gamma_l(P_b)$ is the model \textit{reliability} (in this case, inference correctness probability, or, accuracy), which we define as the probability of issuing a correct inference result under a \gls{ber} $P_b$. %\textcolor{red}{[MF]: my suggestion would be to use a different notation for risk, as it may be confused with the one of data rate. For example, $U_l(P_b)$ denoting inference correctness uncertainty.} 
%For instance, in the case of a classification task, $\Gamma_l(P_b)$ corresponds to the accuracy. 
Instantaneously, the overall reliability depends on the \gls{ai} inference model, the \gls{ber}, and the specific data batch. The instantaneous inference correctness information given a specific input data batch is usually not retrievable during operation, since the ground truth might not be available (e.g., for an object detection or classification task). However, $\Gamma_l(P_b)$ can be estimated a priori on a validation set in the case of a supervised task (the effectiveness of this strategy will be shown in Sec. \ref{sec:numerical_results}). In this paper, we assume the validation set to be drawn from the distribution of the test set, leaving issues related to distribution shifts to future investigations. Otherwise, other metrics, such as the entropy at the output of the classifier can be employed, as in \cite{Merluzzimdpi23}. %In Sec. \ref{sec:performance_eval}, we show the different inference performance of heterogeneous DL models, in the presence of bit-level errors, also in the context of shared spectrum resources. %\textcolor{magenta}{[MF]: I am a bit thoughtful about the term of "trustworthiness" in this context, which is tailored toward security aspects. In our work, incorrect inference output (given a \gls{ber} level) relates to the quality of data transmission and not to a possible attack to the DL model functionality, as we assume that both the \gls{go} user and the AP are not compromised by an adversary. Shall we use the term "robustness to data communication errors" or "\gls{ber}-inference reliability" instead?} \textcolor{blue}{Let's use "reliability" to simplify, I agree that trustworthiness usually refers to something else.}
During slot $t$, the computing delay depends on the selected inference model $l(t)$ and on the computing capacity allocated by the \gls{es}. Denoting the latter by $F(t)$ (measured in Floating Point Operations Per Second - FLOPS), the computing delay reads as $D_{\text{comp}}(t)=\omega_{l(t)}/F(t)$.
Finally, the total delay for $\text{UE}_g$, including communication and computing delays, is $D_{\text{tot}}(t) = D_{\text{tx}}(t) + D_{\text{comp}}(t)$,
%\begin{equation}\label{total_delay}
%    D_{\text{tot}}(t) = D_{\text{tx}}(t) + D_{\text{comp}}(t),
%\end{equation}
where we assume that the inference output transmission in the downlink to be of negligible duration and over a different channel. Finally, the overall delay and the reliability of the employed inference model contribute to the achievement or the failure of the goal, which, in this case, corresponds to \textbf{\textit{correctly classifying data within a deadline}}, as clarified in Sec. \ref{sec:effectiveness}.
%\begin{equation}
%    \overline{A_d}\triangleq\lim_{T\to\infty}\frac{1}{T}\sum\nolimits_{t=0}^{T-1}\mathbb{E}\{A_d(t)\},
%\end{equation} 
\subsubsection{Data-oriented user KPIs}
%For the legacy \gls{do} user, we consider the data rate (in bits/s) as primary KPI. 
As detailed in Sec. \ref{sec:problem_formulation}, the objective of the DO user is to maximize its \textit{average sustained data arrival rate} (i.e., the data arrival rate supported by the system under a set of constraints), while not preventing the \gls{go} user from achieving its target goal performance. This opens new ways of sharing spectrum resources and defining achievable rate regions of interference channels, towards a goal-oriented and compute resource-aware use of the spectrum, as initially suggested in \cite{Merluzzimdpi23}. During a slot $t$ of duration $\tau$, we approximate $\text{UE}_d$ average rate as ~\cite{Filippou15}:
\begin{align}
    R_d(t)=&\frac{1}{\tau}\big[D_{\text{tx}}(t)W\log_2(1+\text{SINR}_d(t))\nonumber \\
    &+ (\tau-D_{\text{tx}}(t))W\log_2(1+\text{SNR}_d(t))\big],
\end{align}
where the first term accounts for the \gls{go} transmission period and the second term for the remaining portion of time\footnote{We assume that one inference data point (or batch) is uploaded by $\text{UE}_g$ per time slot. More involved inference traffic profiles are left for future work.}, with $\text{SNR}_d(t)$ the signal-to-noise ratio obtained by removing the interference term in \eqref{sinr}.
Then, we assume that UE$_d$ generates a continuous flow of data, with new arrivals $A_d(t)$ (in bits) at time $t$ being stored in a buffer before transmission. As clarified in the sequel, inspired by \cite{neely10,Lakshminarayana2014,MerluzziEucnc22}, the goal is to maximize these arrivals while guaranteeing queue stability, considering UE$_d$ as equipped with an infinite-size buffer that evolves as
\begin{equation}\label{ue_d_buffer}
    Q_d(t+1)=\max(0,Q_d(t)-\tau R_d(t))+A_d(t),
\end{equation}
where $\tau$ denotes the slot duration, and $A_d(t)$ the number of arrivals \textit{admitted} to the queue during time slot $t$. We define the sustained data arrival rate as $\bar{A}_d$ (cf. \eqref{long_term_avg}), from which, under the assumption of strong stability (to be guaranteed via the optimization in Sec. \ref{sec:problem_formulation}), by Little's law as in \cite{MerluzziEucnc22}, we can compute the average queuing delay as $\bar{D}_{q,d}=\tau\bar{Q}_d/\bar{A}_d$.
\subsection{Goal-effectiveness and proactive batch drop}\label{sec:effectiveness}
Since both users are served by the same \gls{ap}, we assume the latter to orchestrate resources. Thus, the modulation scheme employed by $\text{UE}_g$ and the wireless channels by the \gls{ap}, which needs to select an \gls{ai} model for inference, allocate computing resources for \gls{go} user data inference, and allow the \gls{do} user to communicate on the same spectrum with a selected transmit power. The latter affects the quality of the received data from the \gls{go} user, and, consequently, the performance of the inference task. Since the transmission delay is known thanks to the knowledge of the modulation order, we assume the overall delay for $\text{UE}_g$ to be known by the AP (or, accurately predictable) at time $t$. 
Then, we assume that a data batch can be proactively dropped if it cannot be treated within a predefined deadline $D_{\max}$, or simply to allow the \gls{do} user to improve its performance. This proactive dropping policy depends on the \textit{altruism} of the UE$_g$ to sacrifice inference quality, with the objective of enhancing UE$_d$ performance. From a protocol perspective, the \gls{ap} takes this decision based on a negotiated service level agreement with the \gls{go} user, with the altruism depending on several factors including, among others, environmental concerns by the GO user, or even more convenient monetary costs agreed with the operator.
To model the drop decision, we denote by $\gamma(t)\in\{0,1\}$ a variable that equals $0$ if the batch is dropped at time $t$. 
Then, we define the \textit{instantaneous} goal-achievement as $\Gamma_{g}(t) = \Gamma_{l(t)}(P_b(t))\cdot \gamma(t)$, with the goal-effectiveness being $\overline{\Gamma}_g$ (cf. \eqref{long_term_avg}).
\begin{definition}[\textbf{\textit{Goal-effective achievable rate region}}]
  Given network conditions (e.g., wireless channels, computing resources, data arrivals), resource allocation, and constraints, it is the set of pairs $(\bar{A}_d,\bar{\Gamma}_g)$ achievable by the system.
\end{definition}
We now propose a problem formulation and solution able to efficiently explore the GEARRs and their boundaries. 
\section{Problem formulation \& solution}\label{sec:problem_formulation}
Maximizing the average data rate of UE$_d$ is a challenging task, due to the variability of data arrivals, wireless channels, and long-term constraints on goal-effectiveness and compute resource usage. We propose a similar approach as proposed in \cite{MerluzziEucnc22}, in which queuing theory and stochastic optimization are exploited to maximize the average throughput of a multi-user network under long-term constraints. First, as introduced in Sec. \ref{sec:sys_model} and illustrated in Fig. \ref{fig:sys_model}, $\text{UE}_d$ is equipped with a buffer to store bits before transmission (cf. \eqref{ue_d_buffer}). The first requirement is for this queue to be strongly stable, i.e., $\bar{Q}_d<\infty$ (cf. \eqref{long_term_avg}).
%\begin{equation}
%    \overline{Q_d}\triangleq\lim_{T\to\infty}\frac{1}{T}\sum\nolimits_{t=0}^{T-1}\mathbb{E}\{Q_d(t)\}<\infty
%\end{equation}
Strong stability is achieved if the departure rate (i.e., $\bar{R}_d$) is greater than the arrival rate (i.e., $\bar{A}_d/\tau$). This condition can be achieved by either increasing the departure rate (i.e., the average UE$_d$ data rate), or decreasing the arrival rate. The former can be increased by increasing \gls{do} user transmit power, and thus interference to the \gls{go} system, while the latter can be achieved via an engineered \textit{proactive} packet/bit drop policy. This results in a fictitious \textit{control valve} (cf. Fig. \ref{fig:sys_model}) that chokes arrivals, thus matching the arrival rate to the \textit{goal-effective capacity} of the system. Therefore, our objective translates into maximizing the arrival rate of $\text{UE}_d$, under long-term constraints on: \textit{i)} its buffer stability, \textit{ii)} a goal-effectiveness threshold for $\text{UE}_g$, \textit{iii)} an average constraint on the number of FLOPS performed by the \gls{es}. In each slot, the optimization variables are: \textit{i)} the data arrivals admitted to the \gls{do} user buffer, \textit{ii)} the $\text{UE}_d$ transmit power $p_d^{\text{tx}}(t)$, iii) the \gls{go} user proactive batch drop, and iv) the allocated computing resources (FLOPS). The long-term problem is formulated as follows:
\begin{align}\label{prob_formulation}
    &\underset{\{\mathbf{\varphi}(t)\}_{\forall t}}{\max}\;\bar{A}_d\\
    &\text{subject to}\quad\textbf{(a)}\; \bar{Q}_d<\infty,\quad\textbf{(b)}\; \bar{\Gamma}_g\geq \bar{\Gamma}_{g,\text{th}}, \quad\textbf{(c)}\; \bar{F}\leq \bar{F}_{\text{th}}\nonumber,\\
    &\textbf{(d)}\; 0\leq A_d(t)\leq A_{d}^{\max}(t),\,\forall t, \quad\textbf{(e)}\; p_{\text{tx},d}(t)\in\mathcal{P},\,\forall t\nonumber,\\
    &\textbf{(f)}\;\gamma\in\{0,1\},\,\forall t,\qquad\qquad\quad\quad\textbf{(g)}\;\gamma(t)D_{\text{tot}}(t)\leq D_{\max},\,\forall t\nonumber,\\
    &\textbf{(h)}\; l(t)\in\mathcal{L},\,\forall t,\qquad\qquad\quad\quad\;\;\, \textbf{(i)}\;0\leq F(t)\leq F_{\max},\,\forall t.\nonumber
\end{align}
Besides long-term constraints $\textbf{(a)}$-$\textbf{(c)}$ on queue stability, goal-effectiveness and average compute resource load, the instantaneous constraints have the following meaning: $\textbf{(d)}$ the admitted arrivals to the queue are non negative and below the actual arrivals at time $t$; $\textbf{(e)}$ the $\text{UE}_d$ transmit power belongs to a predefined discrete set $\mathcal{P}$; $\textbf{(f)}$ a batch is either dropped or transmitted; $\textbf{(g)}$ if transmitted, a batch is treated within the delay threshold; $\textbf{(h)}$ the selected inference model belongs to the set of available models; $\textbf{(i)}$ $F(t)$ the allocated compute resources are non negative and below a maximum value. 

Problem \eqref{prob_formulation} is challenging due to its long-term nature (in terms of objective function and constraints), especially in the absence of a priori statistical knowledge, and non-convexity.\\ %Also, it is non-convex and it involves discrete variables over a long-term horizon.\\
\textbf{Proposed solution.} We propose to solve the problem by transforming \eqref{prob_formulation} into a pure stability problem \cite{neely10}. The latter concerns the buffer, and two \textit{virtual queues} for constraints $(b)$-$(c)$, whose time evolution is respectively defined as follows:
\begin{align}\label{virtual_queues}
&Z(t+1)=\max\left(0,Z(t)-\mu_z\left(\Gamma_g(t)-\bar{\Gamma}_{g,\text{th}}\right)\right)\\
&Y(t+1)=\max\left(0,Y(t)+\mu_y\left(F(t)-\bar{F}_{\text{th}}\right)\right), 
\end{align}
with $\mu_{z(y)}>0$. Following \cite{neely10,Lakshminarayana2014,MerluzziEucnc22}, virtual queues mean rate stability\footnote{For a virtual queue $Z(t)$, it is defined as $\lim_{T\to\infty}\mathbb{E}\{Z(T)\}/T=0$.} is a sufficient condition for guaranteeing the associated constraint, through the definition of the Lyapunov function $L(\mathbf{q(t)}) = \frac{1}{2}Q_d^2(t)+\frac{1}{2}Z^2(t)+\frac{1}{2}Y^2(t)$, with $\mathbf{q(t)}$ denoting a vector that contains all queues (physical and virtual). The mean rate stability is guaranteed by a bounded  \textit{drift-plus-penalty} function, which is defined as follows:
\begin{equation}\label{dpp}
    \delta_p(t)=\mathbb{E}\left\{L(\mathbf{q}(t+1))-L(\mathbf{q}(t))-V A_d(t)|\mathbf{q}(t)\right\},
\end{equation}
with $V$ a trade-off parameter used to balance queue stability (i.e., \gls{do} user delay and constraint violations) and objective function (in this case data arrivals, i.e., \gls{do} user data rate). The higher the value of $V$ is, the closer to optimal the solution is, with a cost on queueing delay for the \gls{do} user. As in \cite{MerluzziEucnc22}, we proceed by instantaneously minimizing an upper bound of \eqref{dpp}, only based on current observation of wireless channels, \gls{go} user modulation scheme, and data arrivals (we omit the derivations due to the lack of space). The problem can be split.
\subsubsection{First sub-problem - optimal data arrivals control}
\begin{equation}\label{sub_arriavls}
    \underset{0\leq A_d(t)\leq A_{d}^{\max}(t)}{\max}\;(V - Q_d(t))A_d(t)
\end{equation}
\eqref{sub_arriavls} is a linear problem that can solved in closed form, and its optimal solution is $A_d^*(t)=A_d^{\max}(t)\cdot\mathbf{1}\{Q_d(t)\leq V\}$. %follows:
%\begin{equation}\label{optim_arrivals}
%    A_d^*(t)=A_d^{\max}(t)\cdot\mathbf{1}\{Q_d(t)\leq V\}
%\end{equation}
\subsubsection{Second sub-problem - $p_{\text{tx},d}(t), \gamma(t),l(t), F(t)$}
The second sub-problem is solved to select $p_{\text{tx},d}(t)$, the inference model $l(t)$, and $F(t)$. It is formulated as follows:
\begin{align}\label{sub_allothers}
    \underset{\{p_{\text{tx},d}, \gamma,l, F\}}{\min}\;&-Q_d(t)\tau R_d(t)-\mu_z Z(t)\cdot\Gamma_g(t)+\mu_yY(t)\cdot F(t)\nonumber\\
    &\text{subject to}\;\textbf{(e)} \text{-}\textbf{(i)}\;\text{of}\;\eqref{prob_formulation}
\end{align}
\begin{figure}
   \centering
   \includegraphics[width=0.95\columnwidth]{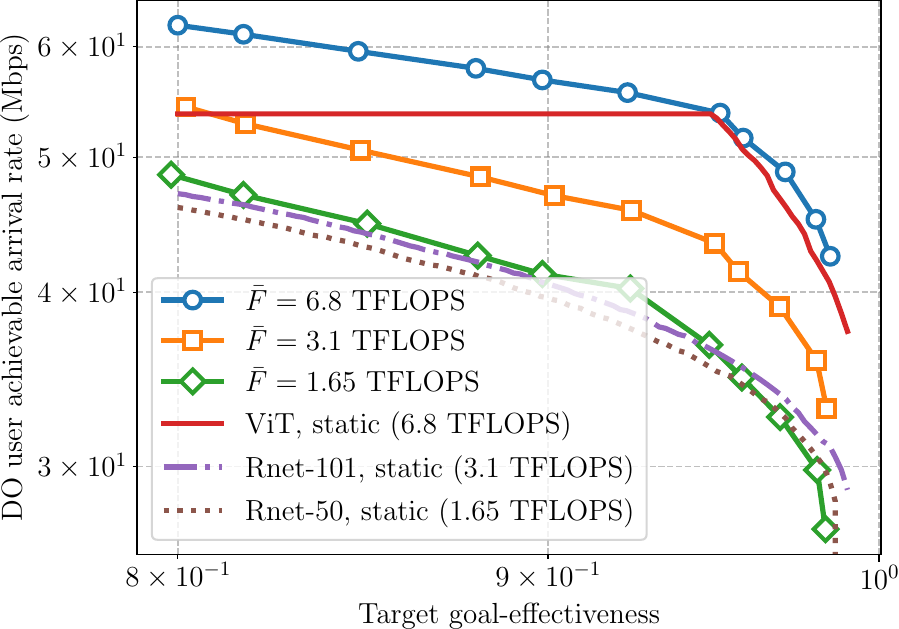}
    \caption{Goal-effective achievable rate regions}
    \label{fig:gearrs_optimized}
    \vspace{-0.5 cm}
\end{figure}
Problem \eqref{sub_allothers} is a mixed-integer non-linear program, however extremely simplified with respect to \eqref{prob_formulation}, as the long-term horizon disappears, and only instantaneous search is needed. Then, assuming a limited number of inference models, we can perform an exhaustive search over $l(t)$, and $p_{\text{tx},d}(t)$, and $\gamma(t)$ to subsequently select $F(t)$. We note that, while it is expected the number of models not to exceed a few units (e.g., due to memory constraints), the presence of more users could make this exhaustive search not scalable. Future investigations will deal with this potential issue via solutions based on multi-agent deep reinforcement learning. %In particular, we can distinguish between two cases, to be compared to take the final decision:\\
%\textbf{Case 1} - $\gamma(t)=0$: in this case, $F^*(t)=0$, $p_{\text{tx},d}^*(t)=p_{\text{tx},d}^{\max}$. Note that this first case is only possible in the \textit{altruistic} regime.\\
%\textbf{Case 2} - $\gamma(t)=1$: in this second case, we need to deal only with the model selection $l(t)$ and $p_{\text{tx},d}(t)$ as discrete variable. As a first step, we perform an exhaustive search over the set of models $\mathcal{L}$ and the set of \gls{do} user transmit power $\mathcal{P}$, 
Concerning the function $\Gamma_{l}(P_b)$ linking reliability to inference model and the \gls{ber}, we estimate it a priori on a validation set. Numerical results will clarify the effectiveness of this procedure, with evaluation on a different test set. Once the model and the transmit power are set, the optimal computing power $F(t)$ is computed as the minimum value guaranteeing constraint $(g)$ of \eqref{prob_formulation}. If the latter cannot be met, the batch is dropped as no timely inference can be performed. This search over the limited set is possible thanks to the decoupling of the problem over time. Then, once resources are optimized, communication occurs for the two users, computation takes place for the \gls{go} user, and all queues are updated (cf. \eqref{ue_d_buffer}, \eqref{virtual_queues}). Finally, the next slot is visited.
\begin{figure*}[t]
    \centering
    \subfloat[Trade-off delay - sustained arrival rate]{
        \includegraphics[width=.3\textwidth]{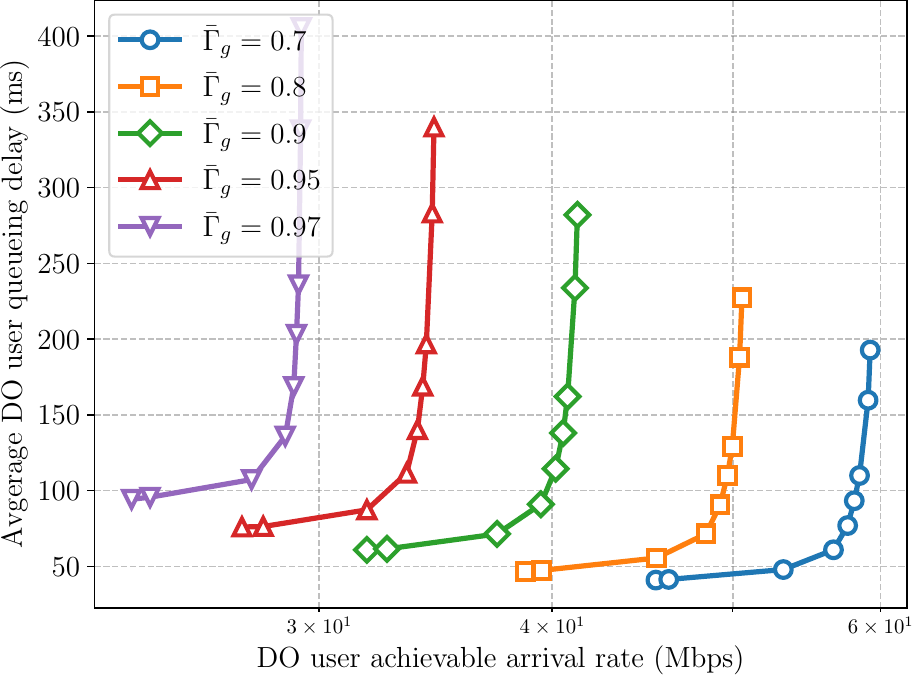}
        \label{fig:delay_vs_rate}
    }
    \subfloat[Evolution of goal-effectiveness]{
        \includegraphics[width=.3\textwidth]{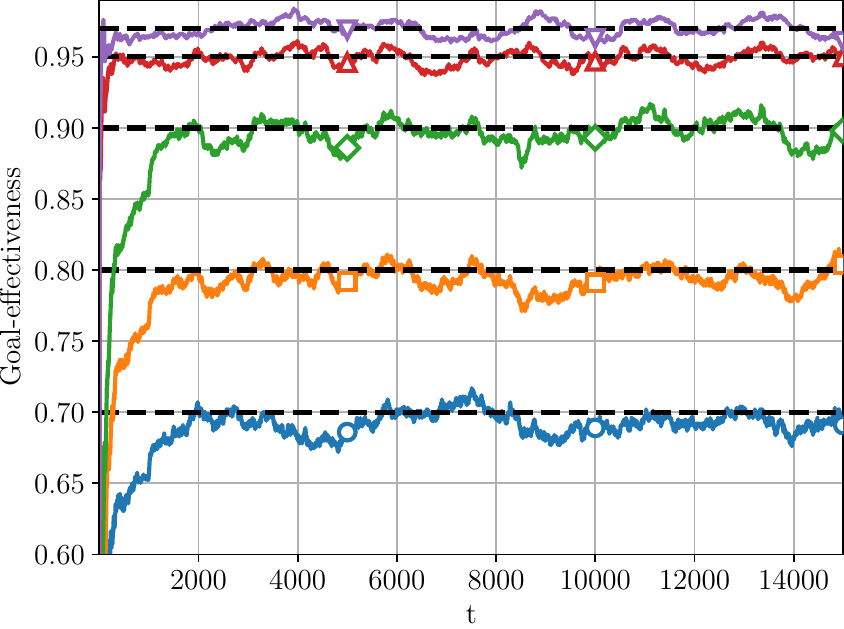}
        \label{fig:measured_effectiveness}
    }
    \subfloat[Evolution of computational load]{
        \includegraphics[width=.3\textwidth]{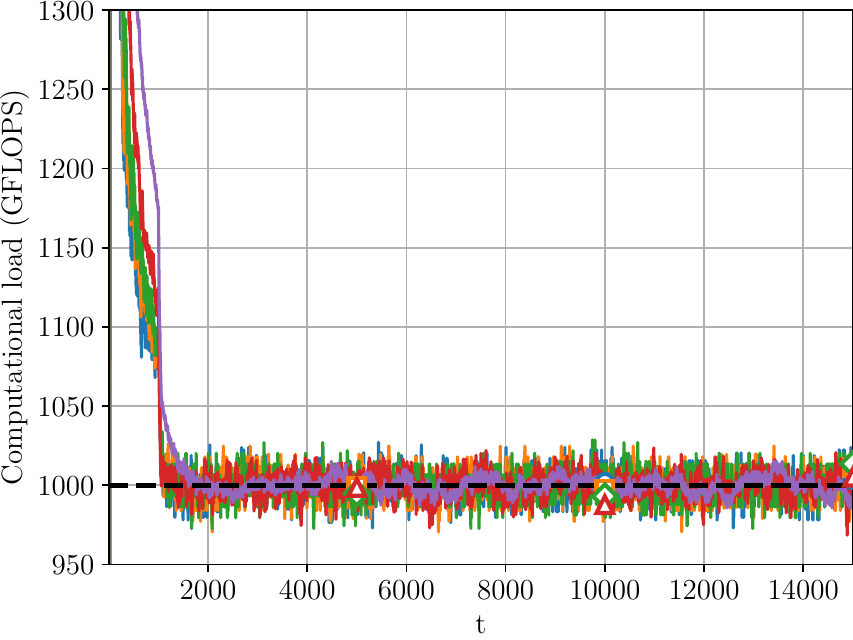}
        \label{fig:measrued_comp_resources}
    }
    \caption{Trade-off between \gls{do} user sustained data rate, goal-effectiveness, and computational load}
    \label{fig:tradeoffs}
    \vspace{-0.5 cm}
\end{figure*}
\section{Numerical evaluation}\label{sec:numerical_results}
\vspace{-0.3 cm}
\begin{table}[h!]
\caption{Simulation parameters}
\centering
\resizebox{\columnwidth}{!}{
\begin{tabular}{|l|l|}
\hline
\textbf{Parameter} & \textbf{Value} \\ \hline
carrier freq. (GHz)/$W$ (MHz)/$\mathcal{M}$ & 3.5/10/256-QAM \\ \hline
noise PSD (dBm/Hz)/noise figure (dB)  & -174/10\\ \hline
channel model/\# of AP antennas & Rician ($K=4$), path loss exp. 3.5/8 \\ \hline
DO/GO/AP positions (x, y) [m] & [-15, 0]/[0, 0]/[0, 20] \\ \hline
DO/GO user transmit power & $p_{\text{tx},d}\in[0,0.1]$ W/$p_{\text{tx},g}=0.1$ W \\ \hline
$D_{\max}$ (ms)/$\tau$ (ms)/$A_{\max}$ (bits) & 20/20/Poisson, $\lambda=5\times10^6$ \\ \hline
$\mathbf{w}_{g(d)}$ (AP combining vector) & Maximum Ratio Combining \\ \hline
\makecell{Inference models \& \\ their comp. load (GFLOPs) \\ Task \& Dataset} & \makecell{Mobilenetv3-small, Resnet-50/101,  vit\_b\_16 \\ 0.11, 8.2/15.6, 33\\ Classification on imagenette \cite{Howard_Imagenette_2019} }\\ \hline
\end{tabular}
}
\vspace{-0.2 cm}
\label{tab:param}
\end{table}
We now show our method's capability of achieving the GEARRs boundaries and exploring the desired trade-offs between \gls{go}, \gls{do} performance, and computational cost. Simulation parameters are reported in Table \ref{tab:param}. Simulations are run for 20000 slots and averaged over the last 10000.

First, we assess the achieved performance of our method in terms of GEARRs, to show its capability to get similar or even better performance compared to a \textit{static} policy (fixed \gls{do} transmit power and inference model), depending on the computational load constraints. For the latter, given a goal-effectiveness constraint, the best data rate is found via an a \textit{posteriori exhaustive search}, and the corresponding computational resources are selected to meet the delay constraint. Then, it should be noted that the \textit{static} policy requires an exhaustive search over the set $\mathcal{P}$ of \gls{do} user transmit power, after statistical parameters are explored. To fairly compare performance, we set constraint $(c)$ in \eqref{prob_formulation} to the values that are needed by the \textit{static} policy to achieve the target goal-effectiveness (computed a posteriori). In Fig. \ref{fig:gearrs_optimized}, we show the GEARRs achieved with our method under the different computational constraints, against the static policy with Resnet-50/101 and vit\_b\_16. As we can notice, the method is able to achieve better performance than the static exhaustive search, with the same respective computational load, with larger gain for lower goal-effectiveness targets. This is thanks to the dynamic decisions based on instantaneous parameters, which allow the system to explore more convenient solutions in the long-term sense (e.g., exploiting favourable channel conditions), and to the proactive \gls{go} dropping policy. These degrees of freedom shrink when imposing higher goal-effectiveness constraints, thus making the method approach the boundaries of the GEARRs that are obtained via the exhaustive search. However, this is achieved in a dynamic way, \textit{only based on instantaneous observations} and without the need to estimate the statistics of the involved variables, which makes the method more suitable for being deployed and work online. This first example shows the capability of our method to achieve the boundaries of the GEARRs by dynamically selecting cross-layer parameters, and guaranteeing all the required long-term constraints.
To further show the flexibility of this framework, in Fig. \ref{fig:tradeoffs}, we show:
\begin{itemize}
    \item[(a)] the average \gls{do} user queueing delay as a function of its achieved arrival rate, under different goal-effectiveness constraints (the different curves), with $\bar{F}_{\text{th}}=1$ TFLOPS;
    \item[(b)] the evolution of the goal-effectiveness, averaged over a $10^3$ samples moving window, for the different thresholds;
    \item[(c)] The evolution of the average computation resources (cf. constraint $\textbf{(i)}$ of \eqref{prob_formulation}) using a $10^3$ samples moving window.
\end{itemize}
From Fig. \ref{fig:delay_vs_rate}, we can notice that, for a given goal effectiveness constraint for the \gls{go} user, the queuing delay at the \gls{do} user increases as a function of the achieved data arrival rate, as predicted by the theory \cite{neely10}, with an asymptotic behaviour around the maximum value (boundary of the GEARR under the imposed constraint). Obviously, a stricter goal-effectiveness requirement results in degraded \gls{do} user performance, showing the complex multi-dimensional trade-off involving different layers. Finally, Figs. \ref{fig:measured_effectiveness} and \ref{fig:measrued_comp_resources} show the convergence of the goal-effectiveness (on the test set using the reliability function estimated on the validation set) and the average computational load toward the desired values.

\section{Conclusion}
We proposed a novel cross-layer and cross-domain resource allocation framework, under which interference-prone spectrum sharing is managed based on higher layer parameters belonging to the world of edge intelligence, namely the diverse model inference capabilities and computational load awareness. Our findings suggest that computing power and inference model robustness to bit-level errors can help boosting the performance of legacy users that use the spectrum to maximize classical metrics such as the data uploading rate. Based on these findings, we proposed a dynamic method that jointly encompasses communication, computing and edge \gls{ai} aspects, toward a computation- and goal-aware spectrum sharing.
\bibliographystyle{IEEEtran}
\bibliography{Main}

\end{document}